\documentstyle[twoside,fleqn,espcrc2]{article}

\newcommand{\Li}[2]{{\mbox{Li}}_{#1}\left(#2\right)}
\newcommand{\Cl}[2]{{\mbox{Cl}}_{#1}\left(#2\right)}
\newcommand{\Ls}[2]{{\mbox{Ls}}_{#1}\left(#2\right)}
\newcommand{\LS}[3]{{\mbox{Ls}}_{#1}^{(#2)}\left(#3\right)}

\newcommand{\be}{\begin{equation}}
\newcommand{\ee}{\end{equation}}
\newcommand{\bea}{\begin{eqnarray}}
\newcommand{\eea}{\end{eqnarray}}
\newcommand{\ep}{\varepsilon}

\newcommand{\nn}{\nonumber}

\renewcommand{\thefootnote}{\fnsymbol{footnote}}

\title{ 
$\qquad\qquad\qquad\qquad\qquad\qquad\qquad\qquad\qquad
 \qquad\qquad\qquad\quad$ MZ-TH/00-22\\[2mm]
Some remarks on the $\ep$-expansion
of dimensionally regulated Feynman diagrams\thanks{Based on the talk
given by A.~D. at the Zeuthen Workshop ``Loops and Legs in Gauge Theories''
(Bastei, Germany, April 2000).}}

\author{A.~I.~Davydychev\address{Department of Physics,
 University of Mainz,
 Staudingerweg 7,
 D-55099 Mainz, Germany}%
\thanks{On leave from
                 Institute for Nuclear Physics, Moscow State University,
                 119899 Moscow, Russia.}
and M.~Yu. Kalmykov$^{\rm a,}$\address{DESY--Zeuthen, 
Theory Group, Platanenallee 6, D-15738 Zeuthen, Germany}%
\thanks{On leave from BLTP, JINR,
                 141980 Dubna, Russia. } }

\begin{document}

\begin{abstract}
Some problems related to construction of the $\ep$-expansion 
of dimensionally regulated Feynman integrals are discussed.
For certain classes of diagrams, an arbitrary term of
the $\ep$-expansion can be expressed in terms of log-sine integrals
related to the polylogarithms. It is shown how the analytic
continuation of these functions can be constructed
in terms of the generalized Nielsen polylogarithms.
\end{abstract}

\maketitle

\renewcommand{\thefootnote}{\arabic{footnote}}
\setcounter{footnote}{0}


{\bf 1.}
Dimensional regularization~\cite{dimreg} is one of
the most powerful tools used in loop calculations.
In some cases, one can derive results valid for an arbitrary
space-time dimension $n$, usually in terms of various 
hypergeometric functions.
However, for practical purposes the coefficients
of the expansion in $\varepsilon$ are important,
where the regulator $\varepsilon$ corresponds to the difference 
between $n$ and the (integer) number of dimensions of interest. 
Below we shall usually imply that $n=4-2\varepsilon$.
In multi-loop calculations higher terms
of the $\varepsilon$-expansion of one- and two-loop functions are needed, 
since they may get multiplied by poles in $\varepsilon$,
not only due to factorizable loops, but also as 
a result of applying the well-known reduction 
techniques \cite{ibp,Tarasov}.

In refs.~\cite{Crete,D-ep}, it was shown that
the log-sine integral functions (see, e.g., in \cite{Lewin},
chapter~7.9),
\begin{equation}
\label{Ls_j}
{\mbox{Ls}}_j(\theta)\equiv -\int\limits_0^{\theta}
\mbox{d}\theta' \ln^{j-1}\left|2\sin\frac{\theta'}{2}\right| ,
\end{equation}  
happen to be very useful to represent results for higher
terms of the $\varepsilon$-expansion. 

For instance, for the one-loop two-point function 
$J^{(2)}(n;\nu_1,\nu_2)$ with 
external momentum $k$, masses $m_1$ and $m_2$ 
and unit powers of propagators,
the following result for an arbitrary term
of the $\varepsilon$-expansion has been obtained
in \cite{Crete,D-ep}:
\begin{eqnarray}
\label{2pt_res2}
&& \hspace*{-7mm}
J^{(2)}(4\!-\!2\varepsilon;1,1) = \mbox{i}\pi^{2-\varepsilon}
\frac{\Gamma(1+\varepsilon)}{2(1-2\varepsilon)}
\nonumber \\
&& \hspace*{-5mm} \times
\bigg\{ \frac{m_1^{-2\varepsilon} \!+\! m_2^{-2\varepsilon}}{\varepsilon}
+ \frac{m_1^2\!-\!m_2^2}{\varepsilon \; k^2}
\left( m_1^{-2\varepsilon} \!-\! m_2^{-2\varepsilon} \right)
\nonumber \\ 
&& \hspace*{-5mm}
+ \frac{\left[\Delta(m_1^2,m_2^2,k^2)\right]^{1/2-\varepsilon}}
       {(k^2)^{1-\varepsilon}}
\sum_{j=0}^{\infty} \frac{(2\varepsilon)^j}{j!}
\nonumber \\  
&& \hspace*{-5mm} \times
\sum_{i=1}^{2}
\left[ {\mbox{Ls}}_{j+1}(\pi) -
{\mbox{Ls}}_{j+1}(2\tau'_{0i})
\right] \bigg\} ,
\end{eqnarray}
where
\begin{eqnarray}
\label{two-point}
&& \cos\tau'_{01} = (m_1^2-m_2^2+k^2)/(2m_1 \sqrt{k^2}) \; , 
\nonumber \\
&& \cos\tau'_{02} = (m_2^2-m_1^2+k^2)/(2m_2 \sqrt{k^2}) \; ,
\nonumber \\ 
&& \cos\tau_{12} = (m_1^2+m_2^2-k^2)/(2m_1 m_2) \; ,
\end{eqnarray} 
whereas the ``triangle'' function $\Delta$ is defined as
\be
\label{Delta}
\Delta(x,y,z) = 2xy+2yz+2zx-x^2-y^2-z^2 .
\ee
One can see that $\tau_{12}+\tau'_{01}+\tau'_{02}=\pi$.
In fact, these angles can be associated with a triangle
whose sides are $m_1$, $m_2$ and $\sqrt{k^2}$.
Moreover, the area of this triangle is 
$\textstyle{1\over4}\sqrt{\Delta(m_1^2,m_2^2,k_{12}^2)}$.
For details of geometrical description, see in \cite{DD}.

Note that the values of ${\mbox{Ls}}_j(\pi)$ can be expressed in terms of
Riemann's $\zeta$ function, see Eqs.~(7.112)--(7.113) of \cite{Lewin}.
The infinite sum with ${\mbox{Ls}}_j(\pi)$ in (\ref{two-point})
can be converted into $\Gamma$ functions,
\begin{equation}
\label{phi=pi/2}
\sum_{j=0}^{\infty} \frac{(2\varepsilon)^j}{j!}\;
\mbox{Ls}_{j+1}(\pi) =
-\pi\; \frac{\Gamma(1+2\varepsilon)}{\Gamma^2(1+\varepsilon)} \; .
\end{equation}

\vspace{3mm}

{\bf 2.} The $\varepsilon$-expansion (\ref{2pt_res2}) is directly applicable
in the region where $\Delta(m_1^2,m_2^2,k^2)\geq 0$, i.e.\ when
$(m_1-m_2)^2\leq k^2\leq(m_1+m_2)^2$. In other regions,
the proper analytic continuation of the occurring $\Ls{j}{\theta}$ 
should be constructed.
To do this, it is convenient to introduce the variable
(cf.\ in ref.~\cite{DT2})
\begin{equation}
\label{def_z}
z \equiv e^{ {\rm i} \sigma \theta }, \hspace{5mm}
\ln(-z) = \ln(z) - {\rm i} \sigma \pi,
\end{equation}
where the choice of the sign $\sigma=\pm 1$ is related to the
causal ``+i0'' prescription for the propagators.

Since ${\mbox{Ls}}_1(\theta)=-\theta$, we get
\begin{equation}
{\rm i} \sigma \left[ \Ls{1}{\pi}  - \Ls{1}{\theta} \right]
= \ln (-z) .
\label{Ls1}
\end{equation}
For the next order, we can use the fact that
${\mbox{Ls}}_2(\theta)={\mbox{Cl}}_2(\theta)$,
where (see in \cite{Lewin})
\begin{eqnarray}
\label{Cl_j}
{\mbox{Cl}}_j(\theta)=\left\{
\begin{array}{l}
\!\!\! {\textstyle{1\over2{\rm i}}}
\left[ {\mbox{Li}}_j\left(e^{{\rm i}\theta}\right)
     \!-\! {\mbox{Li}}_j\left(e^{-{\rm i}\theta}\right)
\right] ,
\hspace{1mm} j \; \mbox{even} \\[2mm]
\!\! {\textstyle{1\over2}}
\left[ {\mbox{Li}}_j\left(e^{{\rm i}\theta}\right)
     \!+\! {\mbox{Li}}_j\left(e^{-{\rm i}\theta}\right)
\right] ,
\hspace{1mm} j \; \mbox{odd} 
\end{array}\!\!\!
\right.
\end{eqnarray}
is the Clausen function, whereas
$\mbox{Li}_j$ is the polylogarithm.
In other words, ${\mbox{Cl}}_j(\theta)$ corresponds either to the
imaginary part or to the real part of
${\mbox{Li}}_j\left(e^{{\rm i}\theta}\right)$,
depending on whether $j$ is even or odd.
Therefore, the analytic continuation reads
\begin{equation}
{\rm i}  \sigma \left[ \Ls{2}{\pi} \! -\! \Ls{2}{\theta} \right]  =
- {\textstyle{1\over2}} \left[ \Li{2}{z} \!-\! \Li{2}{1/z} \right ] ,
\label{Ls2}
\end{equation}
where $ \Ls{2}{\pi} = 0$.
The result for the $\varepsilon$-term
of the two-point function was obtained in \cite{NMB}.

To proceed further, we need similar relations between 
higher $\Ls{j}{\theta}$ and the imaginary (or real) parts 
of the polylogarithms. 
For $j=3$, ${\mbox{Ls}}_3(\theta)$ can be expressed in terms of
the imaginary part of ${\mbox{Li}}_3\left(1-e^{{\rm i}\theta}\right)$.
Then, the imaginary part
of ${\mbox{Li}}_4\left(1-e^{{\rm i}\theta}\right)$ is already
a mixture of ${\mbox{Ls}}_4(\theta)$ and ${\mbox{Cl}}_4(\theta)$,
whereas its real part involves the generalized log-sine integral 
$\mbox{Ls}_4^{(1)}(\theta)$. All these relations can be found
in~\cite{Lewin} (some misprints are mentioned in \cite{D-ep}).
However, attempts to generalize these results to higher functions
show that the relations get more and more cumbersome.

Instead of going that way, we suggest to consider how the
higher $\mbox{Ls}_j$ functions are generated by the
imaginary (and real) parts of the generalized Nielsen polylogarithms 
(see, e.g., in \cite{Nielsen}),
\begin{equation}
\label{Sab}
S_{a,b}(z) = \frac{(-1)^{a+b-1}}{(a-1)! \; b!} \!\int\limits_0^1
\mbox{d} \xi\; \frac{\ln^{a-1}\!\xi \ln^b (1\!-\!z\xi)}{\xi},
\end{equation}
where $S_{a,1}(z) = \Li{a+1}{z}.$
We obtain
\begin{eqnarray}
&& \hspace*{-7mm}
\mbox{Re}\; S_{1,2} (e^{{\rm i} \theta}) = 
{\textstyle{1\over2}} \Cl{3}{\theta} 
+ {\textstyle{1\over2}} \zeta_3 
- {\textstyle{1\over2}} \left( \pi \!-\!\theta \right) \Ls{2}{\theta} , 
\nonumber \\[2mm] && \hspace*{-7mm}
\mbox{Im}\; S_{1,2} (e^{{\rm i} \theta}) = 
- {\textstyle{1\over2}} \Ls{3}{\theta} 
- {\textstyle{1\over24}} \theta  \left( \theta^2  \!-\! 3 \pi \theta 
       \!+\!  3  \pi^2 \right), 
\nonumber
\end{eqnarray}
\begin{eqnarray}
&& \hspace*{-7mm}
\mbox{Re}\; S_{1,3} (e^{{\rm i} \theta}) = 
- {\textstyle{1\over4}} \LS{4}{1}{\theta}
+ {\textstyle{1\over4}} \pi \Ls{3}{\theta}
\nonumber \\ && \hspace*{-2mm}
+ {\textstyle{1\over90}} \pi^4 
+ {\textstyle{1\over48}} \pi^3 \theta 
- {\textstyle{1\over32}} \pi^2 \theta^2 
+  {\textstyle{1\over48}} \pi \theta^3   
- {\textstyle{1\over192}} \theta^4, 
\nonumber \\[2mm] && \hspace*{-7mm}
\mbox{Im}\; S_{1,3} (e^{{\rm i} \theta}) = 
{\textstyle{1\over6}} \Ls{4}{\theta}
+ {\textstyle{1\over4}} \Cl{4}{\theta}
- {\textstyle{1\over4}}\pi \zeta_3
\nonumber \\ && \hspace*{-2mm}
+  {\textstyle{1\over4}} \left( \pi - \theta \right) \Cl{3}{\theta} 
- {\textstyle{1\over8}} \left( \pi - \theta \right)^2 \Ls{2}{\theta}, 
\nonumber 
\end{eqnarray}
where the generalized log-sine integral (see in \cite{Lewin})
is defined as
\begin{equation}
\label{Lsjk}
\LS{j}{k}{\theta} =  - \int\limits_0^\theta
\mbox{d}\theta' \theta'^k \ln^{j-k-1}\left|2\sin\frac{\theta'}{2}\right| \; .
\end{equation}
In particular, $\LS{j}{0}{\theta}=\Ls{j}{\theta}$.
Using these relations, we can express the 
$\mbox{Ls}_3$ and $\mbox{Ls}_4$ functions, 
\begin{eqnarray}
&&  \hspace*{-7mm}
{\rm i} \sigma \left[ \Ls{3}{\pi}  - \Ls{3}{\theta} \right]  = 
\nonumber \\ 
&& \hspace*{-2mm}
S_{1,2}(z) - S_{1,2}(1/z) -   {\textstyle{1\over12}} \ln^3  (-z),   
\label{Ls3}
\\ 
&&  \hspace*{-7mm}
{\rm i}  \sigma \left[ \Ls{4}{\pi}  - \Ls{4}{\theta} \right] =
- 3 \left[ S_{1,3}(z) - S_{1,3}(1/z) \right ]
\nonumber \\ 
&& \hspace*{-2mm}
+ {\textstyle{3\over4}}  \left[ \Li{4}{z} - \Li{4}{1/z} \right ]
\nonumber \\ 
&& \hspace*{-2mm}
-  {\textstyle{3\over4}} \left [ \Li{3}{z} + \Li{3}{1/z}  \right ] \ln (-z)
\nonumber \\ && \hspace*{-2mm}
+  {\textstyle{3\over8}} \left [ \Li{2}{z} - \Li{2}{1/z}  \right ] \ln^2 (-z) \; ,
\label{Ls4}
\end{eqnarray}
where
$\Ls{3}{\pi} = - \frac{1}{12} \pi^3$,
$\Ls{4}{\pi} = \frac{3}{2} \pi \zeta_3$.

We have also constructed further expressions,
\begin{eqnarray}
&& \hspace*{-7mm}
{\rm i}  \sigma \left[ \Ls{5}{\pi}  - \Ls{5}{\theta} \right]  = 
12 \left[ S_{1,4}(z) - S_{1,4}(1/z) \right ]
\nonumber \\ && \hspace*{-2mm}
- 3 \left[ S_{2,3}(z) - S_{2,3}(1/z) \right ]
\nonumber \\ && \hspace*{-2mm}
+ 3 \left[ S_{1,3}(z) + S_{1,3}(1/z) \right ] \ln (-z)
\nonumber \\ && \hspace*{-2mm}
+  {\textstyle{1\over80}} \ln^5  (-z), 
\label{Ls5}
\\ 
&& \hspace*{-7mm}
{\rm i} \sigma \left[ \Ls{6}{\pi}  - \Ls{6}{\theta} \right]  = 
-60 \left[ S_{1,5}(z) - S_{1,5}(1/z) \right ]
\nonumber \\ && \hspace*{-2mm}
+ 15 \left[ S_{2,4}(z) - S_{2,4}(1/z) \right ]
\nonumber \\ && \hspace*{-2mm}
-  {\textstyle{15\over4}} \left[ \Li{6}{z} - \Li{6}{1/z} \right ]
\nonumber \\ && \hspace*{-2mm}
- 15  \left[ S_{1,4}(z) + S_{1,4}(1/z) \right ] \ln (-z)
\nonumber \\ && \hspace*{-2mm}
+ {\textstyle{15\over4}} \left[ \Li{5}{z} + \Li{5}{1/z} \right ] \ln (-z)
\nonumber \\ && \hspace*{-2mm}
- {\textstyle{15\over8}} \left[ \Li{4}{z} - \Li{4}{1/z} \right ] \ln^2 (-z)
\nonumber \\ && \hspace*{-2mm}
+ {\textstyle{5\over8}} \left[ \Li{3}{z} + \Li{3}{1/z} \right ] \ln^3 (-z)
\nonumber \\ && \hspace*{-2mm}
- {\textstyle{5\over32}} \left[ \Li{2}{z} - \Li{2}{1/z} \right ] \ln^4 (-z) \; ,
\label{Ls6}
\end{eqnarray}
with
$ \Ls{5}{\pi} = -\frac{19}{240} \pi^5$, 
$\; \Ls{6}{\pi} = \frac{45}{2} \pi \zeta_5 + \frac{5}{4} \pi^3 \zeta_3$.

Substituting (\ref{Ls1}), 
(\ref{Ls2}), (\ref{Ls3})--(\ref{Ls6})
into Eq.~(\ref{2pt_res2}) (taking $\theta=\tau'_{01}$ or $\theta=\tau'_{02}$,
denoting the $z$'s from Eq.~(\ref{def_z}) as $z_1$ and $z_2$,
and setting $\sigma=-1$), we arrive at the analytic continuation
of the terms of the $\varepsilon$-expansion, up to order $\varepsilon^5$. 
In fact, we have also obtained results for higher $\mbox{Ls}_j$
functions, up to $j=10$, which allowed us to reach the order $\varepsilon^9$. 

It is instructive to consider the limit $m_2\to 0$. 
Introducing the variables $x= m_1^2/k^2$ and $y=m_2^2/k^2$
(and remembering that $\sigma = -1$), we get
\begin{eqnarray*}
&& \hspace*{-7mm}
z_1 \to \frac{y}{(1-x)^2} \!+\! {\cal O}(y^2),
\hspace{3mm}
z_2 \to x  \!+\! \frac{2xy}{1\!-\!x} \!+\! {\cal O}( y^2),
\nonumber \\ && \hspace*{-7mm}
\ln\left[\Delta(m_1^2,m_2^2,k^2)/(k^2)^2\right]
\to - \mbox{i} \pi +  2 \ln (1-x)
\nonumber \\ && \hspace*{20mm}
- \frac{2 y (1+x)}{(1-x)^2}
+ {\cal O}( y^2).
\end{eqnarray*}
Then,
$S_{a,b}(1/z_1)$ can be converted into $S_{a,b}(z_1)$ by means of 
known relations given in Ref.~\cite{Nielsen}. After this, 
the limit $y\to 0$ ($m_2\to 0$) can be taken, since
all $\ln y$ terms are cancelled.
The obtained expressions can be simplified by transforming
$S_{a,b}(z_2)$ into $S_{a,b}(1/z_2)$. In such a way, we also avoid
appearance of terms like $S_{a,b}(-1)$.
Note that for this limit the terms up to order $\varepsilon^3$ can be extracted 
from Eq.~(A.3) of ref.~\cite{FJTV}. Our expressions are in agreement with 
their results.

In fact, using hypergeometric representation
\begin{eqnarray}
\label{BD}
&& \hspace{-7mm}
\left. J^{(2)}(4\!-\!2\varepsilon;1,1)
\right|_{\begin{array}{l} \!\!{}_{m_1=0} \\ \!\!{}^{m_2\equiv m} \end{array}} =
\mbox{i}\pi^{2-\varepsilon} m^{-2\ep}
\frac{\Gamma(1+\ep)}{\ep(1-\ep)}
\nn \\[-3mm]  && \hspace*{20mm} \left.
\times _2F_1\left( \begin{array}{c} 1, \; \ep \\ 
                 2-\ep \end{array} \right| \frac{k^2}{m^2} \right) 
\end{eqnarray}
(see, e.g., Eq.~(10) of \cite{BD-TMF}), 
an arbitrary term of the $\varepsilon$-expansion
can be obtained. Employing Kummer's relations for 
contiguous functions, one can transform the 
$_2F_1$ function from Eq.~(\ref{BD}) into
\[
\frac{1-\ep}{1\!-\!2\ep}
\left\{ \frac{1+u}{2u} - \frac{(1\!-\!u)^2}{2u} 
\left.
_2F_1\left( \!\! \begin{array}{c} 1, \; 1\!+\!\ep \! \\
                 1-\ep \end{array} \right| u \right) 
\right\} \; ,
\]
with $u\equiv k^2/m^2$. 
This (transformed) $_2F_1$ function can be expressed in terms of a simple 
one-fold parametric integral,
\[
(1-u)^{-1-2\ep} \left\{ 1
- \ep \int\limits_0^1 \frac{{\mbox{d}}t}{t}\; t^{-\ep}\; 
\left[ (1\!-\!ut)^{2\ep} \!-\! 1 \right] \right\} \; .
\]
Expanding the integrand in $\ep$, we arrive at 
\begin{eqnarray}
\label{m2=0}
&& \hspace{-8mm}
\left. J^{(2)}(4\!-\!2\varepsilon;1,1)
\right|_{\begin{array}{l} \!\!{}_{m_1=0} \\ \!\!{}^{m_2\equiv m} \end{array}} = 
\mbox{i}\pi^{2-\varepsilon} m^{-2\ep}
\frac{\Gamma(1+\varepsilon)}{(1-2\varepsilon)}
\nonumber \\[-2mm] && \hspace*{-6mm} 
\times
\Biggl\{ \frac{1}{\ep} - \frac{1-u}{2u\ep} 
\left[ (1-u)^{-2\ep} - 1 \right]
\nonumber \\ && \hspace*{-6mm} 
- \frac{(1\!-\!u)^{1\!-\!2\ep}}{u} 
\sum_{j=1}^\infty \ep^j \sum_{k=1}^{j} (-2)^{j-k}
S_{k,j-k+1}(u)
\Biggr\} ,
\end{eqnarray}
which agrees with the results discussed earlier.

\vspace{3mm}

{\bf 3.} 
In ref.~\cite{D-ep} it was shown that similar explicit results can be
constructed for the off-shell massless one-loop three-point function  
with external momenta $p_1$, $p_2$ and $p_3$ ($p_1+p_2+p_3=0$),
\begin{eqnarray}
\label{defJ}
&& \hspace*{-7mm}
J (n; \; \nu_1  ,\nu_2  ,\nu_3 | p_1^2, p_2^2, p_3^2)
\nonumber \\
&& \hspace*{-5mm}
\equiv \int
 \frac{\mbox{d}^n r}{ \left[(p_2 -r )^2\right]^{\nu_1}
\left[(p_1 +r )^2\right]^{\nu_2}
      (r^2)^{\nu_3} } ,
\end{eqnarray}
as well as for the two-loop vacuum diagram with arbitrary masses
$m_1$, $m_2$ and $m_3$,
\begin{eqnarray}
\label{defI}
&& \hspace*{-7mm}
I(n; \; \nu_1, \nu_2, \nu_3 | m_1^2, m_2^2, m_3^2)
\nonumber \\
&& \hspace*{-7mm}
\equiv  \!\!\int\!\!\int\!\! \frac{\mbox{d}^n p \;\; \mbox{d}^n q}
      {\left( p^2 \!-\! m_1^2 \right)^{\nu_1}
       \left( q^2 \!-\! m_2^2 \right)^{\nu_2}
       \left[ (p\!-\!q)^2 \!-\! m_3^2 \right]^{\nu_3}} .
\end{eqnarray}
Results for general $n$ and $\nu_i$ (in terms of hypergeometric functions
of two variables) are available in Refs.~\cite{triangle,BD-TMF,DT1}.
According to the magic connection \cite{DT2}, these integrals 
are closely related to each other. For example,
in the case $\nu_1=\nu_2=\nu_3=1$ this connection 
(see Eq.~(16) of \cite{DT2}) yields
\begin{eqnarray}
J(4-2\varepsilon;1,1,1) = \pi^{-3\varepsilon}\;
\mbox{i}^{1+2\varepsilon}\;
\left( p_1^2 p_2^2 p_3^2 \right)^{-\varepsilon}\;
\nonumber \\
\times
\frac{\Gamma(1+\varepsilon)}{\Gamma(1-2\varepsilon)}\;
I(2+2\varepsilon;1,1,1) ,
\end{eqnarray}
where we assume that $p_i^2\leftrightarrow m_i^2$.
Below we shall omit the arguments $p_i^2$
and $m_i^2$ in the integrals $J$ and $I$, respectively.

Then, using exact results in terms of $_2F_1$ functions
\cite{DT1,FJJ,D-ep}, in combination with the formula 
\begin{eqnarray}
\label{sum_Ls}
\sum_{j=0}^{\infty}\frac{(-2\varepsilon)^j}{j!} {\mbox{Ls}}_{j+1}(2\phi)
= -2\pi\; \frac{\Gamma(1\!-\!2\varepsilon)}{\Gamma^2(1\!-\!\varepsilon)}\;
\theta(-\cos\phi) \;
\hspace*{-8mm}
\nonumber \\
-\frac{2^{1-2\varepsilon}\tan\phi}
      {(1-2\varepsilon)\sin^{2\varepsilon}\phi}\;
\left. _2F_1\left( \!\!\begin{array}{c} 1, 1/2 \\ 3/2\!-\!\varepsilon \end{array}
\!\! \right| -\tan^2\phi \right) ,
\end{eqnarray}
the following results have been obtained in \cite{D-ep}:
\begin{eqnarray}
\label{conjecture}
&& \hspace*{-7mm}
J(4-2\varepsilon;1,1,1) = 2 \pi^{2-\varepsilon}\; 
\mbox{i}^{1+2\varepsilon}\;
\frac{\Gamma(1\!+\!\varepsilon)\Gamma^2(1\!-\!\varepsilon)}
     {\Gamma(1-2\varepsilon)}
\nonumber \\
&& \hspace*{-5mm} \times
\frac{\left[ \Delta(p_1^2,p_2^2,p_3^2)\right]^{-1/2+\varepsilon}}
     {(p_1^2 p_2^2 p_3^2)^{\varepsilon}}
\sum_{j=0}^{\infty} \frac{(-2\varepsilon)^j}{(j\!+\!1)!}
\nonumber \\
&& \hspace*{-5mm}
\times \biggl[ {\mbox{Ls}}_{j+2}(\pi) - \sum_{i=1}^{3}
\big[ {\mbox{Ls}}_{j+2}(\pi) \!-\! {\mbox{Ls}}_{j+2}(2\phi_i) \big] 
\biggr] ,
\end{eqnarray}
\begin{eqnarray}
\label{conjecture2}
&& \hspace*{-7mm}
I(4-2\varepsilon;1,1,1) = \pi^{4-2\varepsilon}\; 
\frac{\Gamma^2(1+\varepsilon)}{(1-\varepsilon)(1-2\varepsilon)}\;
\nonumber \\
&& \hspace*{-5mm} \times
\bigg\{ -\frac{1}{2\varepsilon^2}
\bigg[ \frac{m_1^2+m_2^2-m_3^2}{(m_1^2 m_2^2)^{\varepsilon}}
\nonumber \\
&& \hspace*{-5mm}
+\frac{m_2^2+m_3^2-m_1^2}{(m_2^2 m_3^2)^{\varepsilon}}
+\frac{m_3^2+m_1^2-m_2^2}{(m_3^2 m_1^2)^{\varepsilon}}
\bigg] \; 
\nonumber \\
&& \hspace*{-5mm}
+ \left[\Delta(m_1^2,m_2^2,m_3^2)\right]^{1/2-\varepsilon} 
\sum_{j=0}^{\infty} \frac{(2\varepsilon)^j}{(j\!+\!1)!}
\nonumber \\
&& \hspace*{-5mm}
\times \!\biggl[ {\mbox{Ls}}_{j+2}(\pi) \!-\!\! \sum_{i=1}^{3}
\big[ {\mbox{Ls}}_{j+2}(\pi) \!-\! {\mbox{Ls}}_{j+2}(2\phi_i) \big]
\biggr] \!
\bigg\} ,
\end{eqnarray}
where the angles 
$\phi_i$ ($i=1,2,3$) are defined via
\begin{eqnarray}
&& \cos\phi_1=(p_2^2+p_3^2-p_1^2)/(2\sqrt{p_2^2 p_3^2}),
\nonumber \\
&& \cos\phi_2=(p_3^2+p_1^2-p_2^2)/(2\sqrt{p_3^2 p_1^2}),
\nonumber \\
&& \cos\phi_3=(p_1^2+p_2^2-p_3^2)/(2\sqrt{p_1^2 p_2^2})
\end{eqnarray}
(remember that $p_i^2\leftrightarrow m_i^2$ for the integrals $I$),
so that $\phi_1+\phi_2+\phi_3=\pi$.
Note that the angles $\theta_i$ from \cite{DT1,DT2} are related
to $\phi_i$ as $\theta_i=2\phi_i$.
By analogy with the two-point case (\ref{two-point}), the angles $\phi_i$
can be understood as the angles of a triangle
whose sides are $\sqrt{p_1^2}$, $\sqrt{p_2^2}$ and $\sqrt{p_3^2}$,
whereas its area is $\textstyle{1\over4}\sqrt{\Delta(p_1^2,p_2^2,p_3^2)}$.

For the two lowest orders
($\varepsilon^0$ and $\varepsilon^1$), we reproduce
eqs.~(9)--(10) from \cite{DT2}.
Useful representations for the $\varepsilon^0$ terms of both types
of diagrams can also be found in \cite{ep0}.
Moreover, in Eq.~(26) of \cite{UD3} a one-fold integral representation
for $J(4-2\varepsilon;1,1,1)$ is presented
(for its generalization, see Eq.~(7) of \cite{DT2}). 
Expanding the integrand in $\varepsilon$, we were able to confirm the
$\varepsilon$-expansion (\ref{conjecture}) numerically.

To construct the analytic continuation of the terms of the 
$\varepsilon$-expansion (\ref{conjecture}) 
and (\ref{conjecture2}), we need just to apply 
substitutions  (\ref{Ls1}), (\ref{Ls2}), (\ref{Ls3})--(\ref{Ls6}),
with $\theta=2\phi_i$ ($i=1,2,3$).
The remaining terms $\Ls{j+2}{\pi}$ can actually be treated
in the same way, if we substitute $\theta=0$. In any case, their
values are known (in terms of $\zeta$ function)
and can be summed into a combination of $\Gamma$ functions,
Eq.~(\ref{phi=pi/2}).  
We get three variables $z_i=e^{2{\rm i}\sigma\phi_i}$, see Eq.~(\ref{def_z}),
such that $z_1 z_2 z_3=1$.
The causal prescription requires to take $\sigma=+1$ for 
the $J$-integrals and $\sigma=-1$ for the $I$-integrals.
Using the substitutions presented above, we obtain 
the analytic continuation of the results (\ref{conjecture})
and (\ref{conjecture2}) up to $\varepsilon^4$.
We note that the result for the $\varepsilon$-term 
was known \cite{UD3,DT2} (in terms of $\mbox{Li}_3$).

When the masses are equal, $m_1=m_2=m_3\equiv m$
(this also applies to the symmetric case $p_1^2=p_2^2=p_3^2\equiv p^2$),
the three angles $\phi_i$ are all equal to $\pi/3$,
whereas $\Delta(m^2,m^2,m^2)=3m^4$.
Therefore, in this case the r.h.s. of Eq.~(\ref{conjecture2}) becomes
\begin{eqnarray}
\label{equal_m} 
&& \hspace{-7mm}
\pi^{4-2\varepsilon}\;
\frac{\Gamma^2(1+\varepsilon)\; m^{2-4\varepsilon}}
     {(1-\varepsilon)(1-2\varepsilon)}
\biggl\{
-\frac{3}{2\varepsilon^2}
+ \frac{\sqrt{3}}{3^{\varepsilon}}
\sum_{j=0}^{\infty}
\frac{(2\varepsilon)^j}{(j\!+\!1)!}
\nonumber \\
&& \hspace{-2mm} \times
\left[ 3{\mbox{Ls}}_{j+2}\left({\textstyle{2\pi\over3}}\right)
\!-\!2{\mbox{Ls}}_{j+2}(\pi) \right]
\!\biggr\} .
\end{eqnarray}
For instance, in the contribution of order $\varepsilon$
the transcendental constant ${\mbox{Ls}}_3(2\pi/3)$ appears.
This constant 
was discussed in detail in \cite{DT2}.
The fact that ${\mbox{Ls}}_3(2\pi/3)$ occurs in certain two-loop
on-shell integrals and three-loop vacuum integrals 
has been noticed in \cite{FKK,FKnew,ChS}.
Moreover, in \cite{FKnew} it was observed that the higher-$j$
terms from (\ref{equal_m}) form a basis for certain on-shell 
integrals with a single mass parameter.
Connection of ${\mbox{Ls}}_3(2\pi/3)$ with multiple binomial
sums is discussed in \cite{sum2}. 
We also note that in \cite{KL} the constant $\mbox{Ls}_3(\pi/2)$
appeared. 

\vspace{3mm}

{\bf 4.}
One of the interesting problems is 
to construct terms of the $\varepsilon$-expansion for the one-loop
three-point function with general masses.
In this sense, the geometrical description seems to be
rather instructive. The geometrical approach to the
three-point function is discussed in section~V of \cite{DD}
(see also in \cite{Crete}). This function can be represented
as an integral over a spherical (or hyperbolic)
triangle, as shown in Fig.~6 of \cite{DD},
with a weight factor $1/\cos^{1-2\ep}\theta$
(see eqs.~(3.38)--(3.39) of \cite{DD}).
This triangle 123 is split into three triangles 012, 023
and 031. Then, each of them is split into two
rectangular triangles, according to Fig.~9 of \cite{DD}.
We consider
the contribution of one of the six resulting triangles,
namely the left rectangular triangle in Fig.~9.
Its angle at the vertex~0 is denoted as
${\textstyle{1\over2}}\varphi_{12}^{+}$, whereas the
height dropped from the vertex~0 is denoted $\eta_{12}$.

The remaining angular integration is (see eq.~(5.16) of \cite{DD})
\bea
\label{last}
&& \hspace{-7mm}
\frac{1}{2\ep} \int\limits_0^{\varphi_{12}^{+}/2} \mbox{d}\varphi
\left[ 1 -
\left( 1 + \frac{\tan^2\eta_{12}}{\cos^2\varphi} \right)^{-\ep} \right]
\nn \\
&& \hspace{-7mm}
= {\textstyle{1\over2}} \sum\limits_{j=0}^{\infty}\frac{(-\ep)^j}{(j+1)!} \!
\int\limits_0^{\varphi_{12}^{+}/2} \!\! \mbox{d}\varphi 
\ln^{j+1}\left( 1\! +\! \frac{\tan^2\eta_{12}}{\cos^2\varphi} \right) .
\eea

First of all, we note that the l.h.s. of Eq.~(\ref{last}) yields 
a representation valid for 
an arbitrary $\varepsilon$ (i.e., in any dimension).
To get the result for the general three-point function,
we need to consider a sum of six such integrals.
The resulting representation is closely related
to the representation in terms of hypergeometric functions
of two arguments \cite{Tar-Bastei} (see also in \cite{SRSL} for
some special cases).

In the limit $\ep\to0$ we get a combination
of $\mbox{Cl}_2$ functions, eq.~(5.17) of \cite{DD}.
Collecting the results for all six triangles,
we get the result for the three-point function with
arbitrary masses and external momenta, corresponding
(at $\ep=0$) to the analytic continuation of the
well-known formula presented in \cite{`tHV-79}.   
The higher terms of the $\ep$-expansion correspond to the
angular integrals on the r.h.s. of Eq.~(\ref{last}). 
The problem of constructing closed representations for these
terms, as well as their analytic continuation, 
is very important.
We note that the $\ep$-term of the three-point function with general
masses has been calculated in \cite{NMB} in terms of
$\mbox{Li}_3$. 

\vspace{5mm}

{\bf 5.}
We have shown that the compact structure
of the coefficients of the $\varepsilon$-expansion
of the two-point function (\ref{two-point}), 
the massless off-shell three-point function (\ref{conjecture})
and two-loop massive vacuum diagrams (\ref{conjecture2}),
in terms of log-sine integrals,
allows to perform analytic continuation in terms
of generalized Nielsen polylogarithms (\ref{Sab}),
in some cases (\ref{m2=0}) even for an arbitrary order of the 
$\ep$-expansion.
It is likely that a further generalization of these results 
is possible, e.g.\ for the three-point function with different masses,
two-point integrals with two (and more) loops and three-loop vacuum integrals. 
In particular, 
numerical analysis of the coefficients of the expansion of
certain two-point on-shell integrals and three-loop vacuum integrals
\cite{FKnew} shows that
in some cases the values of generalized log-sine integrals 
$\mbox{Ls}_j^{(k)}$, Eq.~(\ref{Lsjk}), may be involved.
For instance, in ref.~\cite{FKnew} it was shown that 
$\LS{4}{1}{2\pi/3}$ is connected with $V_{3,1}$ from \cite{Bro99}.

The fact that the generalization of $\mbox{Ls}_2=\mbox{Cl}_2$
goes in the $\mbox{Ls}_j$ direction, rather than in $\mbox{Cl}_j$
direction (see Eq.~(\ref{Cl_j})),
is very interesting. There is another example \cite{UD2,Bro93},
the off-shell massless ladder three- and four-point diagrams with
an arbitrary number of loops, when such a generalization went
in the $\mbox{Cl}_j$ direction (for details, see \cite{Crete}). 
It could be also noted that the the two-loop non-planar (crossed) 
three-point diagram gives in this case the square of 
the one-loop function, $({\mbox{Cl}}_2(\theta))^2$ 
(cf.\ Eq.~(23) of \cite{UD3}),
leading to the structure $({\mbox{Cl}}_2(\pi/3))^2$ in the symmetric
($p_i^2=p^2$) case. Recently, these constants have been also found
in massive three-loop calculations \cite{rho,Bro99,GKP}.

The construction of analytic continuation of 
the generalized log-sine functions should be investigated in
more detail. 
In fact, it may require including some other generalizations of
polylogarithms (see, e.g., in Ref.~\cite{RV}).

\vspace{2mm}

{\bf Acknowledgements.}
A.~D. would like to thank the organizers of `Loops and Legs 2000',
it was really a very useful conference.
A.~D.'s research and participation in the conference were supported
by DFG.
M.~K. is grateful to the THEP group (University of Mainz) for their
hospitality during his research stay, which was supported by BMBF 
under contract 05~HT9UMB~4.
At an earlier stage (before November 1999), A.~D.'s research
was supported by the Alexander von Humboldt Foundation;
also partial support from the grants RFBR No.~98--02--16981
and Volkswagen No.~I/73611 is acknowledged.



\begin{thebibliography}{99}

\bibitem{dimreg}
G.~'tHooft and M.~Veltman,
  Nucl. Phys. B44 (1972) 189;\\
C.G.~Bollini and J.J.~Giambiagi,
  Nuovo Cimento   12B (1972) 20;\\
J.F.~Ashmore,
  Lett. Nuovo Cim.   4 (1972) 289;\\
G.M.~Cicuta and E.~Montaldi,
  Lett. Nuovo Cim.   4 (1972) 329.

\bibitem{ibp}
F.V.~Tkachov, Phys. Lett.   100B (1981) 65;\\   
K.G.~Chetyrkin and F.V.~Tkachov,
   Nucl. Phys.   B192 (1981) 159.

\bibitem{Tarasov}
O.V.~Tarasov, Phys. Rev.   D54 (1996) 6479;\\
Nucl. Phys.   B502 (1997) 455.

\bibitem{Crete}
A.I.~Davydychev, Mainz preprint MZ-TH/99-30 (hep-th/9908032).

\bibitem{D-ep} A.I.~Davydychev, Phys. Rev.   D61 (2000) 087701.

\bibitem{Lewin}
L.~Lewin, {\em Polylogarithms and associated functions}
      (North Holland, Amsterdam, 1981).

\bibitem{DD}
A.I.~Davydychev and R.~Delbourgo,
   J. Math. Phys. 39 (1998) 4299.

\bibitem{DT2}
A.I.~Davydychev and J.B.~Tausk,
   Phys. Rev.  D53 (1996) 7381 (hep-ph/9504431).

\bibitem{NMB}
U.~Nierste, D.~M\"uller and M.~B\"ohm, Z. Phys. C57 (1993) 605.

\bibitem{Nielsen}
K.S.~K\"olbig, J.A.~Mignaco and E.~Remiddi, B.I.T.10 (1970) 38;\\
R.~Barbieri, J.A.~Mignaco and E.~Remiddi, Nuovo Cimento A   11 (1972) 824;\\
A.~Devoto and D.W.~Duke, Riv. Nuovo Cim.   7, No.6 (1984) 1.

\bibitem{FJTV}
J.~Fleischer, F.~Jegerlehner, O.V.~Tarasov and O.L. Veretin,
   Nucl. Phys.   B539 (1999) 671.

\bibitem{BD-TMF}
E.E.~Boos and A.I.~Davydychev,
Teor. Mat. Fiz.   89 (1991) 56 [Theor. Math. Phys.   89 (1991) 1052].

\bibitem{triangle}
E.E.~Boos and A.I. Davydychev,
   Vestn. Mosk. Univ. (Ser.3)   28 (No.3) (1987) 8;\\ 
A.I. Davydychev, J. Phys.   A25 (1992) 5587.

\bibitem{DT1}
A.I.~Davydychev and J.B.~Tausk,
   Nucl. Phys. B397 (1993) 123.

\bibitem{FJJ}
C.~Ford, I.~Jack and D.R.T.~Jones,
   Nucl. Phys. B387 (1992) 373.

\bibitem{ep0}
J.S.~Ball and T.-W.~Chiu, Phys. Rev.   D22 (1980) 2550;
          D23 (1981) 3085(E);\\
J.J.~van der Bij and M.~Veltman, Nucl. Phys.   B231 (1984) 205;\\
H.J.~Lu and C.A.~Perez, report SLAC-PUB-5809 (1992);\\
N.I.~Ussyukina and A.I.~Davydychev, Phys. Lett.   B298 (1993) 363.

\bibitem{UD3}
N.I.~Ussyukina and A.I.~Davydychev,  
   Phys. Lett.   B332 (1994) 159.

\bibitem{FKK}
J.~Fleischer, M.Yu.~Kalmykov and A.V.~Ko\-ti\-kov,
   Phys. Lett.   B462 (1999) 169;   B467 (1999) 310(E);\\
J.~Fleischer and M.Yu.~Kalmykov, hep-ph/9907431.

\bibitem{FKnew}
J.~Fleischer and M.Yu.~Kalmykov,  
Phys. Lett.   B470 (1999) 168.

\bibitem{ChS}
K.G.~Chetyrkin and  M.~Steinhauser,
Nucl. Phys.   B573 (2000) 617.

\bibitem{sum2}
J.M.~Borwein,~D.J.~Broadhurst~and J.~Kamnitzer,  
preprint CECM-99-137 (hep-th/0004153);\\
M.Yu.~Kalmykov and O.~Veretin, preprint DESY-00-062 (hep-th/0004010).

\bibitem{KL}
A.V.~Kotikov and L.N.~Lipatov, preprint DESY-00-059 (hep-ph/0004008).

\bibitem{Tar-Bastei}
O.V.~Tarasov, these proceedings.

\bibitem{SRSL}
L.G. Cabral-Rosetti, M.A. Sanchis-Lozano, hep-ph/9809213.

\bibitem{`tHV-79} G.~'tHooft and M.~Veltman, Nucl. Phys.
               B153 (1979) 365.

\bibitem{Bro99} D.J.~Broadhurst,
   Eur. Phys. J.   C8  (1999) 311.

\bibitem{UD2} N.I.~Ussyukina and A.I.~Davydychev,
   Phys. Lett.   B305 (1993) 136.

\bibitem{Bro93} D.J.~Broadhurst,
   Phys. Lett.   B307 (1993) 132.

\bibitem{rho} L.~Avdeev, J.~Fleischer, S.~Mikhailov and O.~Tarasov,
   Phys. Lett.   B336 (1994) 560;   B349 (1995) 597(E);\\
K.G.~Chetyrkin, J.H.~K\"uhn, M.~Steinhauser,
   Phys. Lett.   B351 (1995) 331.

\bibitem{GKP}
S.~Groote, J.G.~K\"orner and A.A.~Pivovarov,
Phys. Rev.   D60 (1999) 061701.

\bibitem{RV}
E.~Remiddi and J.A.M.~Vermaseren, 
Int. J. Mod. Phys. A15 (2000) 725.

\end{thebibliography}
\end{document}